\documentclass[conference]{IEEEtran}
\ifCLASSINFOpdf
\else
 \usepackage[dvips]{graphicx}
\fi
\usepackage[cmex10]{amsmath}
\usepackage{amsmath,amssymb,amsfonts,latexsym,cite}

\def\bdA{\boldsymbol{A}}
\def\bdB{\boldsymbol{B}}

\def\bdD{\boldsymbol{D}}

\def\bdH{\boldsymbol{H}}
\def\bdI{\boldsymbol{I}}

\def\bdQ{\boldsymbol{Q}}
\def\bdR{\boldsymbol{R}}
\def\bdS{\boldsymbol{S}}

\def\bdU{\boldsymbol{U}}
\def\bdV{\boldsymbol{V}}
\def\bdW{\boldsymbol{W}}
\def\bdX{\boldsymbol{X}}
\def\bdY{\boldsymbol{Y}}
\def\bdZ{\boldsymbol{Z}}

\usepackage{algorithm}
\usepackage{algorithmic}
\usepackage{multirow}
\usepackage{amsmath}
\usepackage{xcolor}

\usepackage{setspace}

\usepackage{algorithmic}
\usepackage{array}
\usepackage{mdwmath}
\usepackage{mdwtab}
\usepackage{eqparbox}
\usepackage[tight,footnotesize]{subfigure}
\usepackage{fixltx2e}
\usepackage{stfloats}
\begin{document}
%
% paper title
% can use linebreaks \\ within to get better formatting as desired
\title{One-sided Precoder Designs on Manifolds \\for Interference Alignment }

% author names and affiliations
% use a multiple column layout for up to three different
% affiliations
\author{\IEEEauthorblockN{Chen~Zhang,~ Huarui~Yin, ~and Guo~Wei}
\IEEEauthorblockA{Department of Electrical Engineering and Information Science\\
University of Science and Technology of China, HeFei, Anhui, 230027, P.R.China\\
Email: zhangzc@mail.ustc.edu.cn,~\{yhr,~wei\}@ustc.edu.cn}}

% make the title area
\maketitle

\begin{abstract}
%\boldmath
Interference alignment (IA) is a technique recently shown to achieve the maximum degrees of freedom (DoF) of $K$-user interference channel. 
In this paper,  we focus on the precoder designs on manifolds for IA. By  restricting  the optimization  only at the transmitters' side, it will alleviate the  overhead induced by alternation between the forward and reverse links significantly. Firstly a classical steepest descent (SD) algorithm in multi-dimensional complex space is proposed to achieve feasible IA. Then we reform the optimization problem on Stiefel manifold, and propose a novel SD algorithm based on this manifold  with lower dimensions. Moreover, aiming at further reducing the complexity, Grassmann manifold is introduced to derive corresponding algorithm for reaching the perfect IA.   Numerical simulations show that the proposed  algorithms on manifolds have better convergence performance and higher system capacity than previous methods, also achieve the maximum DoF.

\end{abstract}
% IEEEtran.cls defaults to using nonbold math in the Abstract.
% This preserves the distinction between vectors and scalars. However,
% if the conference you are submitting to favors bold math in the abstract,
% then you can use LaTeX's standard command \boldmath at the very start
% of the abstract to achieve this. Many IEEE journals/conferences frown on
% math in the abstract anyway.

% no keywords

% For peer review papers, you can put extra information on the cover
% page as needed:
% \ifCLASSOPTIONpeerreview
% \begin{center} \bfseries EDICS Category: 3-BBND \end{center}
% \fi
%
% For peerreview papers, this IEEEtran command inserts a page break and
% creates the second title. It will be ignored for other modes.
\IEEEpeerreviewmaketitle

\section{Introduction}
Interference alignment (IA) is a technique recently proposed in \cite{IEEEhowto:Jafarkuser},\cite{IEEEhowto:Maddah} and \cite{IEEEhowto:xchannel}, which can achieve much higher wireless networks capacity than previously believed\cite{IEEEhowto:Jafarreflection}. For the $K$-user $M \times M$ multiple-input
multiple-output (MIMO) interference channel, the sum capacity is
\begin{equation}
{C_{sum}} = {{KM} \over 2}\log (1 + SNR) + o(\log (SNR))
\end{equation}
with  achievability of $KM/2$ degrees of freedom (DoF), which is defined as
\begin{equation}
DoF = \mathop {\lim }\limits_{SNR \to \infty } {{{C_{sum}}} \over {\log (SNR)}}
\end{equation}
It means each transmitter-receiver pair is able to communicate with achievability DoF of $M/2$, irrespective of the number of interferers.

 %The main idea of IA is to coordinate transmitting directions in order to force the interference overlapping at each receiver and therefore reserve half of the signaling space interference-free for the desired signals. 

%An iterative algorithm that utilizes the channel reciprocity to achieve interference alignment, by alternating between the forward and reverse communication links in a distributed way, was proposed in\cite{IEEEhowto:Jafarapp}. 
A feasible way to align interference is to design such a precoder  that coordinates transmitting directions, in order to force the interference overlapping at each receiver, and reserve half of the signaling space interference-free  for the desired signals.
Based on the assumption of  channel reciprocity, some previous works such as \cite{IEEEhowto:Jafarapp}  $-$\cite{IEEEhowto:c3}    iteratively optimize both the precoder matrices and interference suppression filters,  by alternating between the forward and reverse links  to achieve interference alignment in a distributed way. 
%An iterative algorithm that utilizes the channel reciprocity was proposed in \cite{IEEEhowto:Jafarapp}. By alternating between the forward and reverse links in a distributed way, it can achieve interference alignment.
However, with the assumption of channel reciprocity, the applicability of these algorithms are restricted within TDD systems only. Moreover, alternation between the forward and reverse links needs tight synchronization and feedback at each node, which may introduce too much overhead when the channel varies quickly. Furthermore, the transmitters and receivers exchange their ``roles" during each iteration of optimization. Thus this scheme is inappropriate for the receivers  with limited computing ability. %Similar to \cite{IEEEhowto:Jafarapp}, some previous works such as \cite{IEEEhowto:Heath} $-$\cite{IEEEhowto:c3} achieve feasible interference alignment by iteratively optimizing both the precoder matrices and interference suppression filters, which means both the transmitters and receivers are active in the iteration. Obviously the drawbacks we previously encountered will be the bottleneck of application in \cite{IEEEhowto:Heath} $-$\cite{IEEEhowto:c3}.

On the other hand,  most of the previous works above  employ traditional
constrained optimization techniques that work in high dimensional complex space. Unavoidably, several  shortcomings are accompanied with the traditional constrained optimization  techniques such as low-converging speed and high-complexity.

%However, generalizing a given optimization
%algorithm on an abstract manifold is only the first step toward the objective
%of this book. Turning the algorithm into an efficient numerical procedure is a
%second step that ultimately justifies or invalidates the first part of the %effort.

To overcome these limitations, in this paper,
% we offer a strategy to limit the optimization only on the transmitters' side. %For the same purpose, an intuitive steepest descent (SD) algorithm is proposed in \cite{IEEEhowto:Ghauch}, which, however defines inner product and gradient direction in inappropriate topologies. 
 we introduce optimization on matrix manifolds into the precoder design for interference alignment, and  limit the optimization only at the transmitters' side. 
%As a mature mathematical theory,  
Optimization algorithms on manifolds consist the merits of lower complexity and better numerical properties. % However, to the best of our knowledge, optimization on manifolds has not yet been widely applied to interference alignment precoder design. 
Firstly for the sake of  comparison,  by employing classical constrained optimization method,  a steepest descent (SD) algorithm in multi-dimensional complex space is provided to design  the precoder of interference alignment.  
Then we reformulate the constrained optimization problem to an unconstrained and non-degraded one on the complex Stiefel manifold with lower dimensions. We locally parameterize the manifold  by  Euclidean projection from the tangent space onto the manifold instead of the traditional method by moving descent step along the geodesic in  \cite{IEEEhowto:Santa} and \cite{IEEEhowto:Edelman}. Thus the SD algorithm on Stiefel manifold is proposed to achieve feasible interference alignment. To further reduce the computation complexity in terms of dimensions of manifold,  we explore  the unitary  invariance property of our cost function, and solve the optimization problem on the complex Grassmann manifold, then present corresponding SD algorithm on Grassmann manifold for interference alignment.

  %Our work is not a simple combination of  techniques from different realms. 
We not only generalize optimization
algorithm on  manifolds, but also  turn the algorithm into an efficient numerical procedure to achieve perfect interference alignment. Moreover by limiting the optimization algorithms performed at the transmitters' side only,  all the three proposed algorithms are transparent at the receivers. Additionally, overhead generated by synchronization and feedback no longer exits since only transmitters participate in the iteration. %And only transmitters participate in the iteration, thus the overhead and other complications generated by alternations between the forward and reverse links will be %which means it bypasses the overhead and other complications generated by alternation between the forward and reverse network. 
Besides, by relaxing the assumption of channel reciprocity, our algorithms are applicable to both TDD and FDD systems. Furthermore, numerical simulation shows that the novel algorithms on manifolds have better convergence performance and higher system capacity than previous methods. Finally, we prove the convergence of the proposed algorithms.

The paper is organized as following. Interference channel is briefly summarized in Section II %A brief description of the system model of K-user wireless MIMO interference channel is
%introduced in Section II, 
and the mathematical model of interference alignment is presented in Section III, followed by the detailed procedures of all three proposed SD algorithms for interference alignment in Section IV. In section V numerical
simulations and corresponding discussion are stated. And the conclusion is given in the last section.%  Finally the work is concluded in section VI.

\textit{Notation}: We use bold uppercase letters for matrices or vectors. ${\bdX^T}$ and ${\bdX^\dag}$ denote the transpose and the conjugate transpose (Hermitian) of the matrix $\bdX$ respectively. Assuming the eigenvalues of a matrix $\bdX$ and their corresponding eigenvectors are sorted in ascending order, $\lambda _X^i$ denotes the ${i^{th}}$ eigenvalue of the matrix $\bdX$. Then $\bdI$ represents the identity matrix. Moreover $tr(\cdot)$ indicates the trace operation. And the Euclidean norm of $\bdX$ is  ${\left\| \bdX \right\|} = \sqrt{ tr({\bdX^\dag}\bdX)}$.  $\left\lfloor \bdX \right\rfloor $ denotes the subspace spanned by the columns of $\bdX$. %$vec(\bdX)$ represents the staked columns of $\bdX$.
 $\mathbb{C}^{{n \times p}}$ represents the $n \times p$ dimensional complex space assuming $n > p$.  
 $\mathbb{R}^+
$ represents positive real number space. 
$\Re \{ \cdot\} $
 and $\Im \{\cdot\}$ denote the real and imaginary parts of a complex quantity, respectively. Finally $\kappa  = \{ 1,...,K\} $ is the set of integers from 1 to $K$.

\section{System model}
\begin{figure}[htp]
\centering
\includegraphics[width=2.9in]{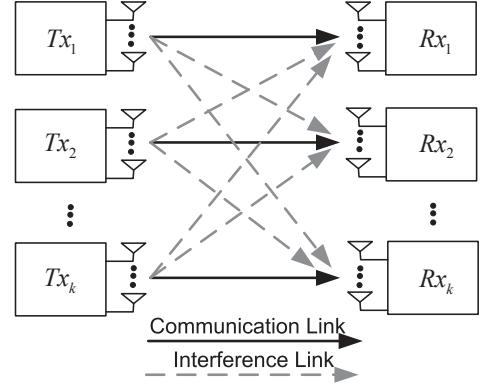}
\caption{K-user MIMO interference channel}
\label{pic1}
\end{figure}

  Consider the K-user wireless MIMO interference channel depicted in Fig. \ref{pic1} where each transmitter and receiver are equipped with ${M^{[k]}}$ and ${N^{[k]}}$ antennas respectively. Each transmitter communicates with its corresponding receiver, and creates interference to all the other receivers.
${d^{[k]}}$ is the desired number of data streams between the ${k^{th}}$   transmitter-receiver pair. Additionally, ${\bdH^{[kj]}}$ denotes the ${N^{[k]}} \times {M^{[j]}}$ channel coefficients matrix from the ${j^{th}}$ transmitter to the ${k^{th}}$ receiver, and is assumed to have i.i.d. complex Gaussian random variables, drawn from a continuous distribution. %We consider a quasi-static channel, therefore the channel realization remains fixed throughout the duration of symbol transmission. 
Finally the received signal vector at receiver $k$ after zero-forcing the interference is denoted by

\begin{equation}
{\overline \bdY ^{[k]}} = {\bdU^{[k]\dag}}{\bdY^{[k]}} = {\bdU^{[k]\dag}}\left(\sum\limits_{j = 1}^K {{\bdH^{[kj]}}{\bdV^{[j]}}} {\bdS^{[j]}} + {\bdW^{[k]}}\right),{\rm{ }}k \in \kappa
\end{equation}
where each element of the ${d^{[j]}} \times 1$ vector ${\bdS^{[j]}}$ represents an independently encoded Gaussian symbol with power ${{{P^{[j]}}} \mathord{\left/
 {\vphantom {{{P^{[j]}}} {{d^{[j]}}}}} \right.
 \kern-\nulldelimiterspace} {{d^{[j]}}}}$ that beamformed with the corresponding ${M^{[j]}} \times {d^{[j]}}$ precoder matrix ${\bdV^{[j]}}$, and then transmitted by the transmitter $j$. ${\bdU^{[k]}}$ is the ${N^{[k]}} \times {d^{[k]}}$ interference zero-forcing filter at the receiver $k$. And ${\bdW^{[k]}}$ is the i.i.d. complex Gaussian noise with zero mean unit variance.

\section{Mathematical Model}
\subsection{Feasibility of Interference Alignment}

The quality of alignment is measured by the interference power
remaining in the intended signal subspace at each receiver.
Therefore interference alignment can be achieved by progressively reducing the power of  leakage interference. And if interference alignment is feasible, the leakage interference can eventually be coordinated to zero. From \cite{IEEEhowto:Jafarapp}, it can be obtained that the ${d^{[k]}}$-dimensional received signal subspace that contains the least interference is the space spanned by the eigenvectors corresponding to the ${d^{[k]}}$-smallest eigenvalues of the interference covariance matrix ${\bdQ^{[k]}}$. Consequently, we try to  minimize the sum of interference power spilled to the desired signal subspaces, by minimizing the sum of the absolute value of ${d^{[k]}}$-smallest eigenvalues of the interference covariance matrix at each receiver  to create ${d^{[k]}}$-dimensional interference-free subspace for the desired signal.

\subsection{Cost Function}
  As previously stated, we try to minimize the sum of the ${d^{[k]}}$-smallest  eigenvalues (in absolute value) of the interference covariance matrix at each receiver, over the set of precoder matrices ${\bdV^{[1]}},...,{\bdV^{[K]}}$ \cite{IEEEhowto:Ghauch}. Therefore, we define the cost function as:
\begin{equation}
\label{object}
\begin{split}
&\mathop {\min }\limits_{{\bdV^{[1]}},...,{\bdV^{[K]}}} f = \sum\limits_{k = 1}^K {} \sum\limits_{i = 1}^{{d^{[k]}}}\left| { {\lambda _{{Q^{[k]}}}^i}} \right| ,~{\rm{ }}k,j \in \kappa\\
&{\rm{subject ~to~  }}{\bdV^{[j]\dag}}{\bdV^{[j]}} = {\bdI_{{d^{[j]}}}}
\end{split}
\end{equation}
where
\begin{equation}
{\bdQ^{[k]}} = \sum\limits_{\scriptstyle j = 1 \hfill \atop
  \scriptstyle j \ne k \hfill} ^K {{{{P^{[j]}}} \over {{d^{[j]}}}}{\bdH^{[kj]}}{\bdV^{[j]}}} {\bdV^{[j]\dag}}{\bdH^{[kj]\dag}}
\end{equation}
is the interference covariance matrix at receiver $k$. With the assumption that all the eigenvalues are sorted in ascending order,  ${\lambda _{{Q^{[k]}}}^i}$ represents the ${i^{th}}$ eigenvalue of the corresponding interference covariance matrix ${{\bdQ^{[k]}}}$. And because ${{\bdQ^{[k]}}}$ is a Hermitian
matrix, all its eigenvalues are real. Therefore,  the cost function $f(\bdV)$, $f:{\mathbb{C}^{n \times p}} \to \mathbb{R}^+ $ is built.

\section{Algorithms on Different Topologies for Interference Alignment}
\subsection{The Steepest Descent Algorithm in Complex Space for IA }
Since our cost function:  $f(\bdV)$, $f:{\mathbb{C}^{n \times p}} \to \mathbb{R}^+ $ is differentiable, intuitively the steepest descent method can be employed to make the cost function converge to a local optimal point efficiently. As well known, the SD method typically can be iteratively processed as: by adding an update increment to the previous iterate in order to
reduce the cost function; the update direction and step size are generally
computed using a local model of the cost function, typically based on 
first  derivatives of the cost function. Therefore we will first find the closed-form expression of the steepest descent direction in ${\mathbb{C}^{n \times p}}$, then employ a property step size rule for each iteration.

As previously stated, the steepest descent method is tightly related to derivative and differentiation. In order to get the derivative of $f(\bdV)$ over $\bdV$, two Jacobian matrices blocks are employed as: 

\begin{footnotesize}
\begin{align}
%\footnotesize
df=
\begin{bmatrix}
\bdD_R^{[1]}&...&\bdD_R^{[K]}
\end{bmatrix}
\begin{bmatrix}
d\bdV_R^{[1]}\\.\\.\\d\bdV_R^{[K]}
\end{bmatrix}+
\begin{bmatrix}
\bdD_I^{[1]}&...&\bdD_I^{[K]}
\end{bmatrix}
\begin{bmatrix}
d\bdV_I^{[1]}\\.\\.\\d\bdV_I^{[K]}
\end{bmatrix}
\end{align}
\end{footnotesize}where $\bdV_R^{[j]} = \Re \{ {\bdV^{[j]}}\}$, and $ \bdV_I^{[j]} = \Im \{ {\bdV^{[j]}}\} $. $\bdD_R^{[j]}$ and $\bdD_I^{[j]}$ are the ${d^{[j]}} \times {M^{[j]}}$ Jacobian matrices which denote the partial differential relation of the cost function over the real and imaginary parts of $\bdV^{[j]}$ respectively. The detail of mathematical derivations can be found in  \cite{IEEEhowto:Ghauch} and \cite{IEEEhowto:Hjorungnes}.
%In\cite{IEEEhowto:Ghauch} the steepest descent direction equals the negative gradient direction which is the definition widely adopted in  multi-dimensional complex space $\mathbb{C}~ {^{n \times p}}$, while the inner-product is, however, defined on the complex Stiefel manifold. 
 Thus, the derivative of $f$ over $\bdV^{[j]}$ is given by
\begin{equation}
\bdD_V^{[j]} = {(\bdD_R^{[j]} + i\bdD_I^{[j]})^T}
\end{equation}

The inner product typically defined in the Euclidean multi-dimensional space is given as:
\begin{equation}
 \left\langle {\bdZ_1},{\bdZ_2} \right\rangle  = tr({\bdZ_2}^\dag {\bdZ_1})
\end{equation}
Then, under the given inner product, the steepest descent direction  is:
\begin{equation}
\label{zsd}
\bdZ^{[j]}=-\bdD_V^{[j]} =- {(\bdD_R^{[j]} + i\bdD_I^{[j]})^T}
\end{equation}
%\begin{equation}
%\dim(\bdZ)=2np
%\end{equation}
Once the formulation of steepest descent direction $\bdZ^{[j]}$ is defined in (\ref{zsd}),  it is necessary to choose a suitable positive step size $\beta^{[j]}$ for each iteration. The Armijo step size rule \cite{IEEEhowto:BoydCVX} states that $\beta^{[j]}$ should be chosen to satisfy the following inequalities:
\begin{equation}
\label{ar1}
f(\bdV) - f(\bdV+\beta \bdZ^{[j]}) \ge {1 \over 2}\beta^{[j]}  \left\langle {\bdZ^{[j]},\bdZ^{[j]}} \right\rangle 
\end{equation}
\begin{equation}
\label{ar2}
f(\bdV) - f(\bdV+2\beta \bdZ^{[j]}) < \beta^{[j]} \left\langle {\bdZ^{[j]},\bdZ^{[j]}} \right\rangle 
\end{equation}
Rule (\ref{ar1}) guarantees that the step $\beta^{[j]}\bdZ^{[j]}$ will expressively decrease the cost function, whereas (\ref{ar2}) undertakes that the step $2\beta^{[j]}\bdZ^{[j]}$ would not be a better choice. A direct procedure for acquiring  a suitable $\beta^{[j]}$ is to keep on doubling $\beta^{[j]}$ until (\ref{ar2}) no longer holds and then halving $\beta^{[j]}$ until it satisfies (\ref{ar1}). 
It can be proved that such $\beta^{[j]}$ can always be found \cite{IEEEhowto:Polak}.

Consolidating all the ideas stated above, we present our  algorithm in \textbf{{Algorithm \ref{alg:SD}}}.
%The structure of the proposed algorithm is explicit.
%Here some explanations are presented. 
In Step 3 and Step 4, the Armijo step rule is performed to find a proper convergence step length. Generally speaking, Step 3 ensures the chosen step ${\beta ^{[j]}}$ will significantly reduce the cost function while Step 4 prevents ${\beta ^{[j]}}$ from being too large that may miss the potential optimal point.
 The operator $gs(\cdot)$ means Gram-Schmidt Orthogonalization \cite{IEEEhowto:zhang} of a matrix, which guarantees the newly computed solution $ {\bdV^{[j]}}$ (or $\bdB_1^{[j]}$, $\bdB_2^{[j]}$ ) still satisfies the unitary
constraint.
%Moreover, the inner product in complex space $\mathbb{C}~ {^{n \times p}}$  is commonly defined as $ < {\bdZ^{[j]}},{\bdZ^{[j]}} >  = tr({\bdZ^{[j]H}}{\bdZ^{[j]}})$.
\algsetup{indent=2em}
\begin{algorithm}[h]
\caption{The Steepest Descent Algorithm in Complex Space for IA}\label{alg:SD}

\begin{algorithmic}

\begin{spacing}{1.5}
\small\STATE Start with arbitrary precoder matrices ${\bdV^{[1]}},...,{\bdV^{[K]}}$, %${\bdV^{[j]\dag}}{\bdV^{[j]}} = \bdI$,
 set initial step size ${\beta ^{[j]}}=1$ and begin iteration.
\STATE for {$j=1...K$}
\STATE (1) Compute the  Jacobian matrix $\bdD_R^{[j]}$ and $\bdD_I^{[j]}$
\STATE (2) Get the steepest descent direction \\\quad\quad\quad\quad\quad ${\bdZ^{[j]}} =-{\bdD_V^{[j]}} =-{(\bdD_R^{[j]} + i\bdD_I^{[j]})^T}$
\STATE (3) Compute $\bdB_1^{[j]} =  gs({\bdV^{[j]}} + 2{\beta ^{[j]}}{\bdZ^{[j]}})$,\\
~~~if $f({\bdV^{[1]}},...,{\bdV^K}) - f({\bdV^{[1]}},...,{\bdV^{[j - 1]}},\bdB_1^{[j]},{\bdV^{[j + 1]}}...{\bdV^{[K]}})  $\\
~~~${\ge\beta ^{[j]}} tr( {\bdZ^{[j]\dag}}{\bdZ^{[j]}} ) $, then set ${\beta ^{[j]}}: = 2{\beta ^{[j]}}$, repeat Step 3.
\STATE (4) Compute $\bdB_2^{[j]} = gs ({\bdV^{[j]}} + {\beta ^{[j]}}{\bdZ^{[j]}})$,\\~~~~if $f({\bdV^{[1]}},...,{\bdV^{[K]}}) - f({\bdV^{[1]}},...,{\bdV^{[j - 1]}},\bdB_2^{[j]},{\bdV^{[j + 1]}}...{\bdV^{[K]}}) $\\~~~$ <{1 \over 2} {\beta ^{[j]}}tr( {\bdZ^{[j]\dag}}{\bdZ^{[j]}} )$, then set ${\beta ^{[j]}}: = {1 \over 2}{\beta ^{[j]}}$, repeat Step 4.
\STATE (5)${\bdV^{[j]}} = gs ({\bdV^{[j]}} + {\beta ^{[j]}}{\bdZ^{[j]}})$
\STATE(6) Continue till the cost function $f$ is sufficiently small.
\end{spacing}

\end{algorithmic}
\end{algorithm}

\textit{Discussion}: 
\begin{itemize}
\item (i) The inner product and the gradient direction are defined in different topologies in\cite{IEEEhowto:Ghauch}. However it is considered to be inappropriate because the gradient is defined after the inner product is given only. In other words, the inner product and the gradient direction must be defined in the same topology. Our proposed algorithm rectifies the topology flaw in \cite{IEEEhowto:Ghauch}, thus avoids the risk of non-convergence.

\item (ii) It can be concluded that Algorithm 1 belongs to the classical optimization method, which means it woks in multi-dimensional space $\mathbb{C}^{n \times p}$ with the dimensions:

\begin{equation}
\label{dimc}
\dim(\mathbb{C}{^{n \times p}})=np
\end{equation}

Obviously the algorithm complexity increases with the dimensions. As discussed before, optimization algorithms on manifolds work in an embedded or quotient space whose dimension  
will be  much smaller than that of classical
constrained optimization methods. Thus, optimization
algorithms on  manifolds  not only have  lower complexity, 
but also perform better numerical properties. The corresponding algorithms on manifolds  will be stated for IA in the next two subsections. 
\end{itemize}
\subsection{The Steepest Descent Algorithm on Complex Stiefel Manifold for IA}
%()Generally speaking, a $d$-dimensional manifold can be informally defined as a set $\mathcal{ M}$ covered
%with a ¡°suitable¡± collection of coordinate patches, or charts, that identify
%certain subsets of $\mathcal{ M}$  with open subsets of $\mathbb{C}^d$ \cite{IEEEhowto:Absil}.% As previously stated, classical
%constrained optimization techniques work in an embedded space whose dimension  
%can be  much larger dimension than that of the manifold. Therefore, optimization
%algorithms that work on the manifold will have  a lower complexity and
%quite often also have better numerical properties.

%a new iterate is generated
%by adding an update increment to the previous iterate in order to
%reduce the cost function. The update direction and step size are generally
%computed using a local model of the cost function, typically based on (approximate)
%first and second derivatives of the cost function, at each step.
Informally, a manifold is a space that is ``modeled on" Euclidean space. It can be defined as  a subset of Euclidean space which is locally the graph of a smooth  function.

Conceptually, the simplest approach to optimize a differentiable function
is to continuously translate a test point in the direction of steepest
descent on the constraint set until one reaches a point where
the gradient is equal to zero. However, there are two challenges for optimization on manifolds.
 First, in
order to define algorithms on manifolds, these operations above must be translated
into the language of differential geometry. Second, once the test point shifts  along the steepest descent direction, it must be retracted back to the manifold. 
Therefore, after reformulating the optimization problem on the Stiefel manifold, we introduce definitions about   project operation and tangent space  for   retraction and gradient respectively.

In many cases, the underlying symmetry
property can be exploited to reformulate the problem as a non-degenerate
optimization problem on  manifolds associated with
the original matrix representation. Thus the constraint condition ${\bdV^{[j]\dag}}{\bdV^{[j]}} = {\bdI}$ in the cost function (\ref{object}), inspires us to solve the problem on the complex Stiefel manifold.  The complex Stiefel manifold \cite{IEEEhowto:zhang} $St(n,p)$ is the set satisfying
\begin{equation}
St(n,p) = \{ \bdX \in \mathbb{C} {^{n \times p}}:{\bdX^\dag}\bdX = \bdI\}
\end{equation}
$St(n,p)$ naturally embeds in $\mathbb{C} {^{n \times p}}$ and inherits the usual topology of  $\mathbb{C} {^{n \times p}}$. It is a compact manifold and from Proposition 3.3.3 in \cite{IEEEhowto:Absil}, we can get:
\begin{equation}
\label{dim}
\dim (St(n,p)) = np - {1 \over 2}p(p + 1)
\end{equation}

Another important definition is the projection. Assuming $\bdY \in {\mathbb{C}^{n \times p}}$ is a rank-$p$ matrix, the projection operator $\pi_{st}(\cdot) :{\mathbb{C}^{n \times p}} \to St(n,p)$ is given by
\begin{equation}
\label{pro}
\pi_{st} (\bdY) = \arg \mathop {\min }\limits_{\bdX \in St(n,p)} {\left\| {\bdY - \bdX} \right\|^2}
\end{equation}
It can be proved that there exits a unique solution if $\bdY$ has full column rank \cite {IEEEhowto:Absil}. From (\ref{pro}), it can be acquired that the projection of an arbitrary rank-$p$ matrix $\bdY$ onto the Stiefel manifold is defined to be the point on the Stiefel manifold closest to $\bdY$ in the Euclidean norm\cite{IEEEhowto:Manton}. Besides, if the singular value decomposition (SVD) of $\bdY$ is $\bdY = \bdU\sum {\bdV^\dag}$, then
\begin{equation}
\pi_{st} (\bdY) = \bdU{\bdI_{n \times p}}{\bdV^\dag}
\end{equation}
based on the Proposition 7 in \cite {IEEEhowto:Manton}.
 
\begin{figure}[htp]
\centering
\includegraphics[width=2.5in]{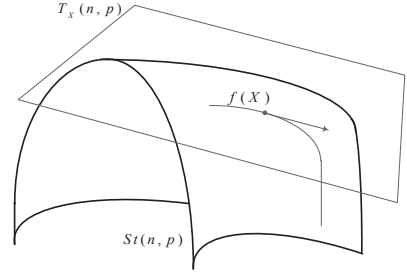}
\caption{Tangent space of Stiefel manifold}
\label{tan}
\end{figure}

Consider $\bdX \in St(n,p)$ and its disturbing point ${\pi _{st}}(\bdX + \varepsilon \bdY) \in St(n,p)$ for certain directions matrix $\bdY \in \mathbb{C}{^{n \times p}} $ and scalar $\varepsilon  \in \mathbb{R}$ . If $\bdY$ satisfies $f({\pi _{st}}(\bdX + \varepsilon \bdY)) = f(\bdX) + O({\varepsilon ^2})$ which means certain directions $\bdY$ do not cause ${\pi _{st}}(\bdX + \varepsilon \bdY)$ to move away from $\bdX$ as $\varepsilon$ increases. The collection of such directions $\bdY$ is called the normal space at $\bdX$ of $St(n,p)$ \cite{IEEEhowto:Absil}. The tangent space  ${T_X}(n,p)$ is defined to be the orthogonal complement of the normal space, which can be roughly illustrated as Fig. \ref{tan}. 
And the mathematical expression of
the tangent space ${T_X}(n,p)$ at $\bdX \in St(n,p)$ is defined by
\begin{equation}
\begin{split}
{T_X}(n,p) = \{ \bdZ \in \mathbb{C}{^{n \times p}}:\bdZ = \bdX\bdA + {\bdX_ \bot }\bdB,\bdA \in \mathbb{C}{^{p \times p}},\\
\bdA + {\bdA^\dag} = 0,\bdB \in \mathbb{C}{^{(n - p) \times p}}\}
\end{split}
\end{equation}
 in which ${\bdX_ \bot } \in {\mathbb{C}^{n \times (n - p)}}$ is defined to be any matrix satisfying ${\left[ {\bdX~{\bdX_ \bot }} \right]^\dag}\left[ {\bdX~{\bdX_ \bot }} \right] = \bdI$ and is the complement of $\bdX\in St(n,p) $. Also from\cite{IEEEhowto:Manton}, it can be obtained that the gradient of our cost function is  in the tangent space ${T_X(n,p)}$. And the dimension of   ${T_X(n,p)}$ is:
\begin{equation}
\label{dim2}
\dim({T_X}(n,p))=p(2n-p)
\end{equation}
  Obviously the steepest descent algorithm requires the computation of the gradient. As we previously emphasize,  the gradient is only defined after ${T_X}(n,p)$
is given an inner product:
\begin{equation}
\label{inner}
\left\langle {\bdZ_1},{\bdZ_2}\right\rangle  = \Re \{ tr(\bdZ_2^\dag(\bdI - {1 \over 2}\bdX{\bdX^\dag}){\bdZ_1})\}
\end{equation}
where ${\bdZ_1},{\bdZ_2} \in {T_X}(n,p)$ and $\bdX \in St(n,p)$. The derivation of (\ref{inner}) can be found in \cite{IEEEhowto:Edelman}. Therefore, under the defined inner product, the steepest descent direction of the cost function $f(\bdX)$
at the point $\bdX \in St(n,p)$ is
\begin{equation}
\label{sd}
\bdZ = \bdX\bdD_X^\dag\bdX - {\bdD_X}
\end{equation}
where $\bdD_X$ is the derivative of $f(\bdX)$. The proof of (\ref{sd}) is in \cite{IEEEhowto:Manton}.

Inspired by \cite{IEEEhowto:Manton} the proposed SD algorithm on complex Stiefel manifold is presented in \textbf{Algorithm \ref{alg:ST}}.  From (\ref{inner}) and (\ref{sd}), it can be easily obtained that the inner product needed for the Armijo step rule is
\begin{equation}
 \left\langle{\bdZ^{[j]}},{\bdZ^{[j]}} \right\rangle  = \Re \{ tr({\bdZ^{[j]\dag}}(\bdI - {1 \over 2}{\bdV^{[j]}}{\bdV^{[j]}}^\dag){\bdZ^{[j]}})\}
\end{equation}
which is used in Step 4 and 5; and the steepest descent on Stiefel manifold of our cost function is
\begin{equation}
{\bdZ^{[j]}} = {\bdV^{[j]}}\bdD_V^{[j]\dag}{\bdV^{[j]}} - \bdD_V^{[j]}
\end{equation}
which is used in Step 3.   Noticing that the project operation $\pi_{st}(\cdot)$ in Step 6 (Step 4, 5) guarantees the newly
computed solution $\bdV^{[j]}$ (or $\bdB_1^{[j]}$, $\bdB_2^{[j]}$ ) after iteration still satisfies ${\bdV^{[j]}} \in St(n,p)$. Using the method of SVD, we can easily compute the project operation.

\algsetup{indent=2em}
\begin{algorithm}[h]
\caption{The Steepest Descent Algorithm on Complex Stiefel Manifold for IA}\label{alg:ST}
\begin{algorithmic}
\begin{spacing}{1.5}
\small\STATE Start with arbitrary precoder matrices ${\bdV^{[1]}},...,{\bdV^{[K]}}$, %${\bdV^{[j]}} \in St({M^{[j]}},{d^{[j]}})$, 
set initial step size ${\beta ^{[j]}}=1$ and begin iteration.
\STATE for {$j=1...K$}
\STATE (1) Compute the  Jacobian matrix $\bdD_R^{[j]}$ and $\bdD_I^{[j]}$
\STATE (2) Then get the derivative of $f$:\\\quad  \quad\quad \quad \quad \quad $\bdD_V^{[j]} = {(\bdD_R^{[j]} + i\bdD_I^{[j]})^T}$
\STATE (3) Get the steepest descent direction \\ \quad \quad \quad  \quad \quad \quad ${\bdZ^{[j]}} = {\bdV^{[j]}}\bdD_V^{[j]\dag}{\bdV^{[j]}} - \bdD_V^{[j]}$
\STATE (4) Compute $\bdB_1^{[j]} = \pi_{st} ({\bdV^{[j]}} + 2{\beta ^{[j]}}{\bdZ^{[j]}})$,\\

~~if $f({\bdV^{[1]}},...,{\bdV^K}) - f({\bdV^{[1]}},...,{\bdV^{[j - 1]}},\bdB_1^{[j]},{\bdV^{[j + 1]}}...{\bdV^{[K]}}) \ge$\\~~$ {\beta ^{[j]}}\Re \{ tr({\bdZ^{[j]\dag}}(\bdI - {1 \over 2}{\bdV^{[j]}}{{\bdV^{[j]\dag}}}){\bdZ^{[j]}})\}$, then set ${\beta ^{[j]}}: = 2{\beta ^{[j]}}$,\\~~and repeat Step 4.
\STATE (5) Compute $\bdB_2^{[j]} = \pi_{st} ({\bdV^{[j]}} + {\beta ^{[j]}}{\bdZ^{[j]}})$,\\~if $f({\bdV^{[1]}},...,{\bdV^{[K]}}) - f({\bdV^{[1]}},...,{\bdV^{[j - 1]}},\bdB_2^{[j]},{\bdV^{[j + 1]}}...{\bdV^{[K]}}) < $\\~~${1 \over 2}{\beta ^{[j]}}\Re \{ tr({\bdZ^{[j]\dag}}(I - {1 \over 2}{\bdV^{[j]}}{{\bdV^{[j]\dag}}}){\bdZ^{[j]}})\}$, then set ${\beta ^{[j]}}: = {1 \over 2}{\beta ^{[j]}}$, \\~~and repeat Step 5.
\STATE (6)${\bdV^{[j]}} = \pi_{st} ({\bdV^{[j]}} + {\beta ^{[j]}}{\bdZ^{[j]}})$
\STATE(7) Continue till the cost function $f$ is sufficiently small.
\end{spacing}
\end{algorithmic}
\end{algorithm}
\textit{Discussion}: 
\begin{itemize}
\item (i)  As previous stated, the algorithms in \cite{IEEEhowto:Santa} and \cite{IEEEhowto:Edelman}  are performed  by  moving the descent step
along the geodesic of  the constrained surface within each iteration. A disadvantage of
this method is the redundant computational cost for
calculating the path of a geodesic \cite{IEEEhowto:Manton}. In this paper, we locally parameterize the manifold  by  Euclidean projection from the tangent space onto the manifold instead of moving along a geodesic,  to achieve a modest reduction  in the computational complexity of the algorithms.
\item (ii)  
Recall (\ref{dimc}) and (\ref{dim2}), it can be obtained that when we reformulate the problem from $\mathbb{C}{^{n \times p}}$ to $St(n,p)$, the dimension of the optimization problem decreases from $np$ to $np - {1 \over 2}p(p + 1)$. Although such dimension-dissension can be observed clearly, we still intend to reduce the dimensions of the space which the optimization algorithm works in. Thus the Grassmann manifold and its corresponding algorithm for IA are stated in the following subsection.
\end{itemize}

\subsection{The Steepest Descent Algorithm on Complex Grassmann Manifold for IA }

Notice that our cost function $f(\bdV)$ satisfies $f(\bdV\bdU) = f(\bdV)$ for any unitary matrix $\bdU$. Because
 \begin{align}
 {\bdQ^{[k]}}(\bdV\bdU)& = \sum\limits_{\scriptstyle j = 1 \hfill \atop 
  \scriptstyle j \ne k \hfill} ^K {{{{P^{[j]}}} \over {{d^{[j]}}}}{\bdH^{[kj]}}{\bdV^{[j]}}{\bdU^{[j]}}} {\bdU^{[j]\dag }}{\bdV^{[j]\dag }}{\bdH^{[kj]\dag }}\nonumber\\
&= \sum\limits_{\scriptstyle j = 1 \hfill \atop 
  \scriptstyle j \ne k \hfill} ^K {{{{P^{[j]}}} \over {{d^{[j]}}}}{\bdH^{[kj]}}{\bdV^{[j]}}\bdI} {\bdV^{[j]\dag }}{\bdH^{[kj]\dag }}\nonumber\\
& = {\bdQ^{[k]}}(\bdV)
\end{align}
which means that multiplying unitary matrix $\bdU$ does not change the eigenvalues and their corresponding eigenvectors of the interference covariance matrix at each receiver. Thus our cost function $f$ should be minimized on the Grassmann manifold rather than on the Stiefel manifold. This is because the Grassmann manifold treats $\bdV$  and $\bdV\bdU$ as equivalent points, leading to a further reduction in the dimension of the optimization problem.
Similar with the previous section, we firstly introduce the definition about Grassmann manifold, then present the project operation and tangent space of Grassmann manifold for retraction and gradient respectively.
% Intuitively, the steepest descent method goes hand-in-hand with derivative and differentiation. Firstly some definitions and derivations about the complex Grassmann manifold are introduced. Then we will derive the steepest decent direction on Grassmann manifold.

The complex Grassmann manifold $Gr(n,p)$ is defined to be the set of all $p$-dimensional complex subspaces of ${\mathbb{C}^{n \times p}}$. Grassmann manifold can be thought as a quotient space of the Stiefel manifold: $Gr(n,p) \simeq St(n,p)/St(p,p)$. Quotient space is more difficult to visualize, as it is not defined as
set of matrices; rather, each point of the quotient space is an equivalence
class of $n\times p$ matrices.

However, we can understand quotient space in this way: assuming $\bdX \in St(n,p)$  is a point on the Stiefel manifold, the columns of $\bdX$ span an orthonormal basis for a $p$-dimensional quotient subspace. That is to say, if $\left\lfloor \bdX \right\rfloor $ denotes the subspace spanned by the columns of $\bdX$, then  $\bdX \in St(n,p)$ implies $\left\lfloor \bdX \right\rfloor  \in Gr(n,p)$. Therefore, there is a one-to-one mapping between points on the Grassmann manifold $Gr(n,p)$ and equivalence classes of $St(n,p)$.  From (\ref{dim}), it can be acquired that:
\begin{align}
\label{dimg}
\dim (Gr(n,p)) &= \dim (St(n,p)) - \dim (St(p,p)) \nonumber\\&= p(n - p)
\end{align}

Let $\bdY \in {\mathbb{C}^{n \times p}}$  be a rank-$p$ matrix. The projection operator $\pi_{gr}(\cdot) :{\mathbb{C}^{n \times p}} \to Gr(n,p)$ onto the Grassmann manifold is defined to be
\begin{equation}
\label{pro2}
\pi_{gr} (\bdY) = \left\lfloor {\arg \mathop {\min }\limits_{\bdX \in St(n,p)} \left\| {\bdY - \bdX} \right\|^2} \right\rfloor 
\end{equation}
It also can be proved that there exits a unique solution if $\bdY$ has full column rank \cite {IEEEhowto:Absil}.
From (\ref{pro2}), it can be acquired that the projection of an arbitrary rank-$p$ matrix $\bdY$ onto the Grassmannn manifold is defined to be the subspace spanned by the point on the Stiefel manifold closest to $\bdY$ in the Euclidean norm. Besides, if the QR decomposition of $\bdY$ is $\bdY=\bdQ\bdR$, the following equality holds:
\begin{equation}
\label{qr2}
\pi_{gr2} (\bdY) = \left\lfloor {\bdQ{\bdI_{n \times p}}} \right\rfloor 
\end{equation}
The proof of (\ref{qr2}) also can be found in\cite{IEEEhowto:Manton}. From (\ref{qr2}), it is obvious that $\pi_{gr}(\bdY)$  is the subspace spanned by the first $p$ columns of $\bdQ$.

 As discussed before, Grassmann manifold is a quotient space of the Stiefel manifold, thus its tangent space is a subspace of the Stiefel manifold's tangent space \cite{IEEEhowto:Absil}. If $\bdX \in St(n,p)$ , the tangent space ${T_{\left\lfloor \bdX \right\rfloor }}(n,p)$ at $\left\lfloor \bdX \right\rfloor  \in Gr(n,p)$  of Grassmann manifold is:
\begin{equation}
{T_{\left\lfloor \bdX \right\rfloor }}(n,p) = \{ \bdZ \in {\mathbb{C}^{n \times p}}:\bdZ = {\bdX_ \bot }\bdB,\bdB \in {\mathbb{C}^{(n - p) \times p}}\} 
\end{equation}
%where ${\bdX_ \bot } \in {\mathbb{C}^{n \times (n - p)}}$ is defined to be any matrix satisfying ${\left[ {\bdX~{\bdX_ \bot }} \right]^\dag}\left[ {\bdX~{\bdX_ \bot }} \right] = \bdI$ and is the complement of $\bdX\in St(n,p) $. 
Recall (\ref{dim2}),  the dimension of the tangent space ${T_{\left\lfloor \bdX \right\rfloor }}(n,p)$ of complex Grassmann manifold is:
\begin{align}
\label{dim4}
\dim{(T_{\left\lfloor \bdX \right\rfloor }}(n,p))&= \dim (T_X(n,p)) - \dim (T_X(p,p)) \nonumber\\&= p(2n-2p)
\end{align}

Besides  the inner product of ${T_{\left\lfloor X \right\rfloor }}(n,p)$ is given by:
\begin{align}
\label{inner2}
\left\langle {\bdZ_1},{\bdZ_2} \right\rangle  = \Re \{ tr({\bdZ_2}^\dag {\bdZ_1})\},~&{\bdZ_1},{\bdZ_2} \in {T_{\left\lfloor \bdX \right\rfloor }}(n,p),\nonumber\\&\bdX \in St(n,p)
\end{align}
The derivation of (\ref{inner2}) can be found in \cite{IEEEhowto:Edelman}.
Therefore, under the defined
inner product, the steepest descent direction \cite{IEEEhowto:Manton} of   the
cost function $f(\bdX)$ at the point $\bdX \in Gr(n,p)$ is:
\begin{equation}
\label{sd2}
\bdZ =  - (\bdI - \bdX{\bdX^\dag}){\bdD_X}
\end{equation}
where $\bdD_X$ is the derivative of $f(\bdX)$.

% Considering our cost function $f(\bdV)$ is $f:{\mathbb{C}~^{n \times p}} \to \Re $, we can get $\bdD_V$ by using two Jacobian matrices blocks:
%
%\begin{footnotesize}
%\begin{align}
%\footnotesize
%df=
%\begin{bmatrix}
%\bdD_R^{[1]}&...&\bdD_R^{[K]}
%\end{bmatrix}
%\begin{bmatrix}
%d\bdV_R^{[1]}\\.\\.\\d\bdV_R^{[K]}
%\end{bmatrix}+
%\begin{bmatrix}
%\bdD_I^{[1]}&...&\bdD_I^{[K]}
%\end{bmatrix}
%\begin{bmatrix}
%d\bdV_I^{[1]}\\.\\.\\d\bdV_I^{[K]}
%\end{bmatrix}
%\end{align}
%\end{footnotesize}
%where $\bdV_R^{[j]} = \Re \{ {\bdV^{[j]}}\}$, and $ \bdV_I^{[j]} = \Im \{ {\bdV^{[j]}}\} $. $\bdD_R^{[j]}$ and $\bdD_I^{[j]}$ are the ${d^{[j]}} \times {M^{[j]}}$ Jacobian matrices which denote the partial differential relation of the cost function over the real and imaginary parts of $\bdV^{[j]}$ respectively. The detail of mathematical derivations can be found in  \cite{IEEEhowto:Ghauch} and \cite{IEEEhowto:zhang}. Thus, the derivative of $f$ over $\bdV^{[j]}$ is given by
%\begin{equation}
%\bdD_V^{[j]} = {(\bdD_R^{[j]} + i\bdD_I^{[j]})^T}
%\end{equation}
\algsetup{indent=2em}
\begin{algorithm}[h]
\caption{The Steepest Descent Algorithm on Complex Grassmann Manifold for IA }\label{alg:GR}
\begin{algorithmic}
\begin{spacing}{1.5}
\small\STATE Start with arbitrary precoder matrices ${\bdV^{[1]}},...,{\bdV^{[K]}}$, %$\bdV^{[j]\dag}\bdV^{[j]}=\bdI$ which implies ${\bdV^{[j]}} \in St({M^{[j]}},{d^{[j]}})$, 
set initial step size ${\beta ^{[j]}}=1$ and begin iteration.
\STATE for {$j=1...K$}
\STATE (1) Compute the  Jacobian matrix $\bdD_R^{[j]}$ and $\bdD_I^{[j]}$
\STATE (2) Then get the derivative of $f$:\\\quad  \quad\quad \quad \quad \quad $\bdD_V^{[j]} = {(\bdD_R^{[j]} + i\bdD_I^{[j]})^T}$
\STATE (3) Get the steepest descent direction \\ \quad \quad \quad  \quad \quad \quad ${\bdZ^{[j]}} =  - (\bdI - {\bdV^{[j]}}{\bdV^{[j]H}})\bdD_V^{[j]}$
\STATE (4) Compute $\bdB_1^{[j]} = \pi_{gr} ({\bdV^{[j]}} + 2{\beta ^{[j]}}{\bdZ^{[j]}})$,\\

~~if $f({\bdV^{[1]}},...,{\bdV^K}) - f({\bdV^{[1]}},...,{\bdV^{[j - 1]}},\bdB_1^{[j]},{\bdV^{[j + 1]}}...{\bdV^{[K]}}) \ge$\\~~${\beta ^{[j]}} tr( {\bdZ^{[j]H}}{\bdZ^{[j]}} )$, then set ${\beta ^{[j]}}: = 2{\beta ^{[j]}}$, and repeat Step 4.
\STATE (5) Compute $\bdB_2^{[j]} = \pi_{gr} ({\bdV^{[j]}} + {\beta ^{[j]}}{\bdZ^{[j]}})$,\\~if $f({\bdV^{[1]}},...,{\bdV^{[K]}}) - f({\bdV^{[1]}},...,{\bdV^{[j - 1]}},\bdB_2^{[j]},{\bdV^{[j + 1]}}...{\bdV^{[K]}}) < $\\~~${1 \over 2} {\beta ^{[j]}}tr( {\bdZ^{[j]H}}{\bdZ^{[j]}} )$, then set ${\beta ^{[j]}}: = {1 \over 2}{\beta ^{[j]}}$, and repeat Step 5.
\STATE (6)${\bdV^{[j]}} = \pi_{gr} ({\bdV^{[j]}} + {\beta ^{[j]}}{\bdZ^{[j]}})$
\STATE(7) Continue till the cost function $f$ is sufficiently small.
\end{spacing}
\end{algorithmic}
\end{algorithm}
The proposed SD algorithm on complex Grassmann manifold is presented in \textbf{Algorithm \ref{alg:GR}}.  Similar with the previous proposed algorithms, the Armijo step rule is performed to find a proper convergence step length.
From (\ref{inner2}) and (\ref{sd2}), it can be easily concluded that the inner product needed for the Armijo step rule is 
\begin{equation}
 \left\langle{\bdZ^{[j]}},{\bdZ^{[j]}} \right\rangle  = tr({\bdZ^{[j]\dag }}{\bdZ^{[j]}})
\end{equation}
which is used in Step 4 and 5; and the steepest descent on Grassmann manifold of our cost function is 
\begin{equation}
{\bdZ^{[j]}} =  - (\bdI - {\bdV^{[j]}}{\bdV^{[j]\dag }})\bdD_V^{[j]}
\end{equation}
which is used in Step 3.   And the project operation $\pi_{gr}(\cdot)$ in Step 6 (Step 4, 5) retracts the newly
computed solution $\bdV^{[j]}$ (or $\bdB_1^{[j]}$, $\bdB_2^{[j]}$) back onto Grassmann manifold $Gr(n,p)$. Using QR decomposition, we can easily compute the project operation.

\begin{figure}[htp]
\centering
\includegraphics[width=2.5in]{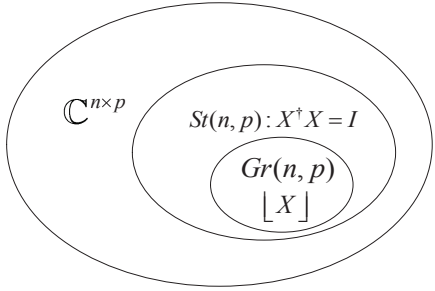}
\caption{Trend of dimension-descension}
\label{c}
\end{figure}

\textit{Discussion}: 
\begin{itemize}

\item Comparing $\dim(Gr(n,p))=p(n-p)$ in (\ref{dimg}) with $\dim(St(n,p))=np-{1 \over 2}p(p+1)$ in (\ref{dim}), a further dimension reduction can be observed. Similarly,  from (\ref{dim4}) we can see  another advantage of using the Grassmann manifold rather than the Stiefel manifold is that ${T_{\left\lfloor \bdX \right\rfloor }}(n,p)$ has only $p(2n-2p)$dimensions, whereas tangent space of $St(n,p)$  has $p(2n-p)$dimensions.
And from \cite{IEEEhowto:Jafarreview}, it can be obtained that in our system model, if each transceiver is equipped with same amount of antenna ($M=N$), then
\begin{equation}
\sum\limits_{k = 1}^K {{d^{[k]}}}  = K \cdot d = {{K \cdot M} \over 2}
\end{equation}
and
\begin{equation}
d={M\over 2}
\end{equation}
Recall (\ref{dimc}), (\ref{dim}) and (\ref{dimg}), we can get that if $M$ is large enough ( $M$ not only can  represent the number of antennas each transceiver equipped, but also can refer to the number of time extension slots \cite {IEEEhowto:Jafarkuser} \cite {IEEEhowto:Jafarreview}), hence
\begin{equation}
\label{b1}
{{\dim (St(M,d))} \over {\dim ({\mathbb{C}~^{M \times d}})}}  \approx 
 {3 \over 4}
\end{equation}
 which is a clear evidence for dimension-descension.
And
\begin{equation}
\label{b2}
{{\dim (Gr(M,d))} \over {\dim ({\mathbb{C}~^{M \times d}})}} = {1 \over 2}
\end{equation}
holds for any integer $M$. (\ref{b2}) means that optimization on Grassmann manifold would reduce dimension further. The trend of dimension-descension  can be roughly illustrated in Fig. \ref{c}.
\end{itemize}

\section{Numerical Results and Discussion}
Without symbol  extension, the feasible condition of k-user interference alignment \cite{IEEEhowto:Jafarreview} is given by:
\begin{equation}
{M^{[j]}} + {N^{[j]}} \ge (K + 1){d^{[j]}}
\end{equation}
For satisfying feasibility and simple computation, we consider a 3-user $2 \times 2$ MIMO interference channel where the desired DoF per user ${d^{[j]}}$ is 1. %Furthermore, in order to compare the convergence performance, within each realization of the simulation, 
All the algorithms are executed under the same scenario including randomly generated channel coefficients, initial precoder matrices and convergence step length. We simulated the proposed three SD algorithms through 100 simulation realizations. 
 
\begin{figure}
\centering
\includegraphics[width=3.5in]{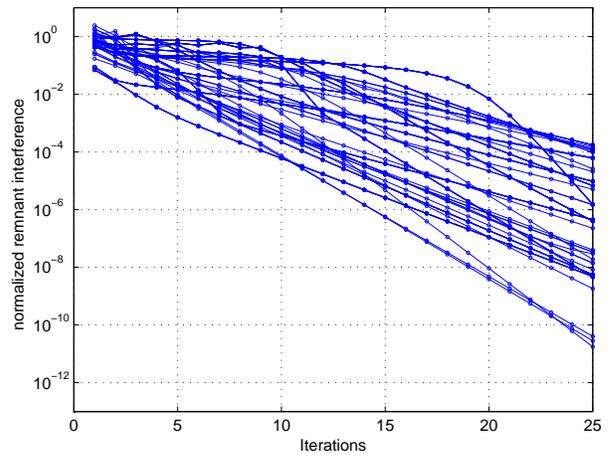}
\caption{Normalized remnant interference of the SD algorithm in complex space}
\label{sdfsd}
\end{figure}

\begin{figure}
\centering
\includegraphics[width=3.5in]{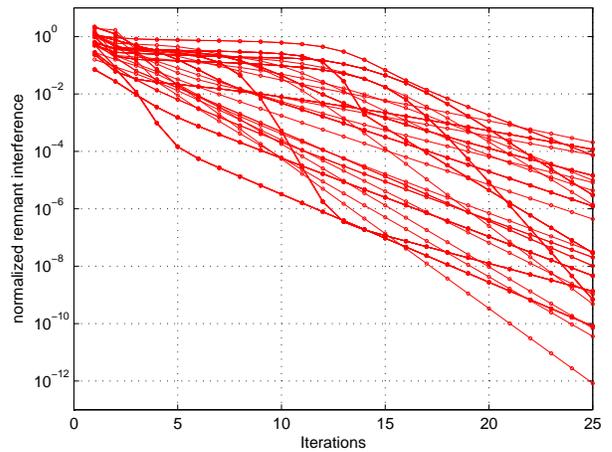}
\caption{Normalized remnant interference of the SD algorithm on  Stiefel manifold}
\label{stfsd}
\end{figure}

\begin{figure}
\centering
\includegraphics[width=3.5in]{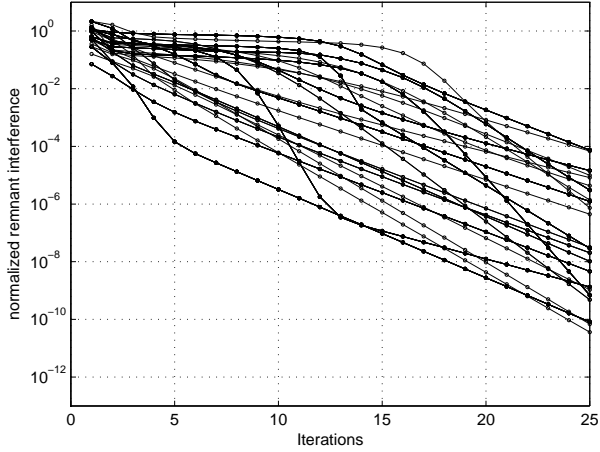}
\caption{Normalized remnant interference of the SD algorithm on  Grassmann manifold}
\label{grfsd}
\end{figure}

\begin{figure}
\centering
\includegraphics[width=3.5in]{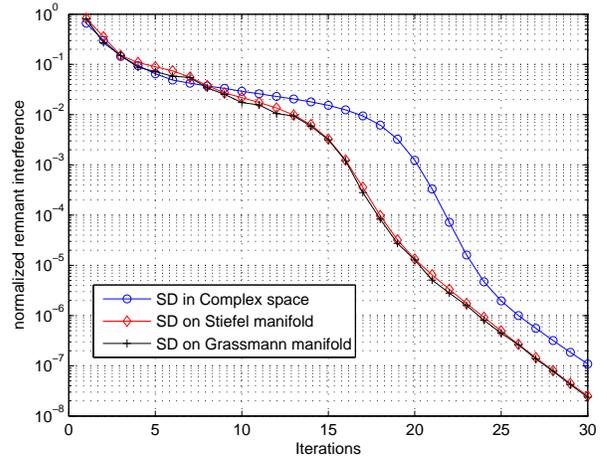}
\caption{Convergence performance}
\label{fsdcompare}
\end{figure}

As shown in Fig. \ref{sdfsd}, Fig. \ref{stfsd} and Fig. \ref{grfsd}, each curve represents an individual simulation realization and all results converge after 20 or more iterations, which is a clear indication of the algorithm convergence performance in such limited iterations.  In order to compare the convergence performance, the average values of 100 realizations results are illustrated in Fig.  \ref{fsdcompare}. It can be observed that the  algorithms on  manifolds  have better convergence performance comparing with the classical optimization method as our expectation.

 This is attributed to the reason that we reformulate the constrained optimization problem to an unconstrained one on  manifolds with lower complexity and better numerical properties; then locally parameterize the manifolds by a Euclidean projection of the tangent space on to the manifolds instead of moving along the geodesic, as stated in the previous sections. Moreover the convergence performance curve of SD algorithm on Stiefel  manifold and the curve of SD algorithm on Grassmann manifold almost overlap. Recall that optimization on Grassmann manifold would reduce dimension further, therefore the SD algorithm on Grassmann manifold will guarantee performance and reduce the computation complexity at the same time.

\begin{figure}
  \centering 
  \subfigure[ SD algorithm in complex space]{ 
    \label{aa} %% label for first subfigure 
    \includegraphics[width=3.5in]{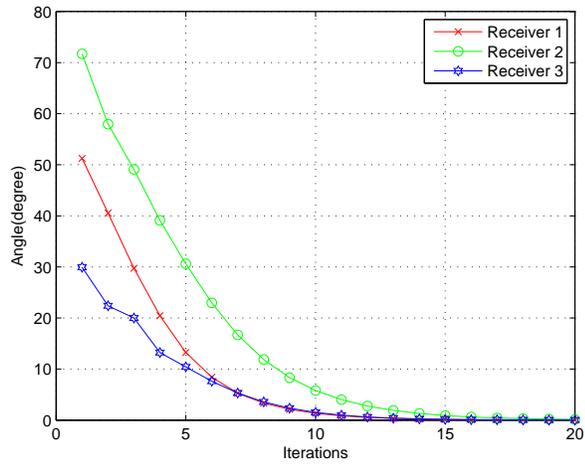}} 
  \hspace{1in} 
  \subfigure[ SD algorithm on Stiefel manifold]{ 
    \label{ab} %% label for second subfigure 
    \includegraphics[width=3.5in]{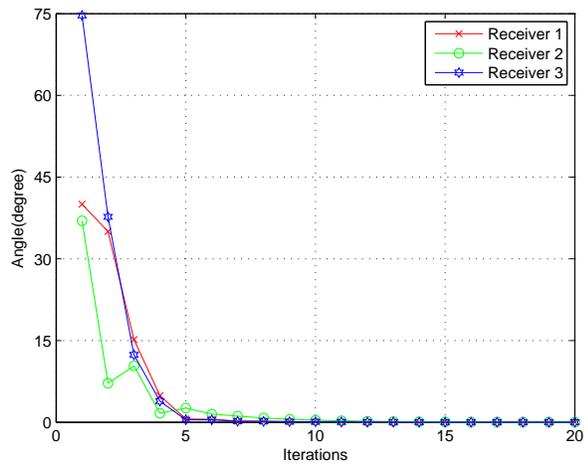}}
	\hspace{1in}  
	\subfigure[ SD algorithm on Grassmann manifold]{ 
    \label{ac} %% label for second subfigure 
    \includegraphics[width=3.5in]{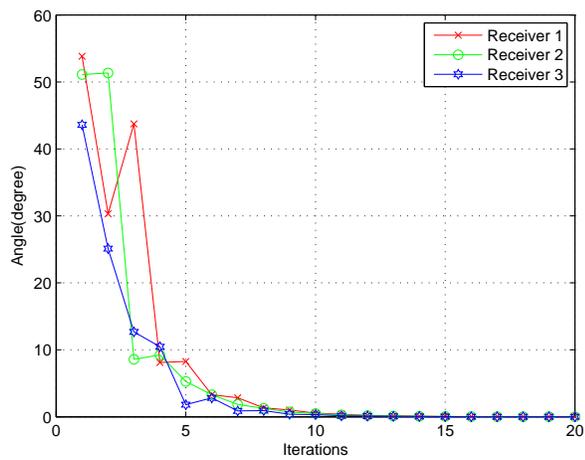}} 
	  \caption{Angles between interfering spaces at each receiver} 
  \label{angle} %% label for entire figure 
\end{figure}

Meanwhile, since there are two interference signals at each receiver, as shown in Fig. \ref{angle}, the angles between the spaces spanned by each interference signals asymptotically converge to zero within one simulation realization, which is another  evidence for achieving the perfect interference alignment.  

\begin{figure}
\centering
\includegraphics[width=3.5in]{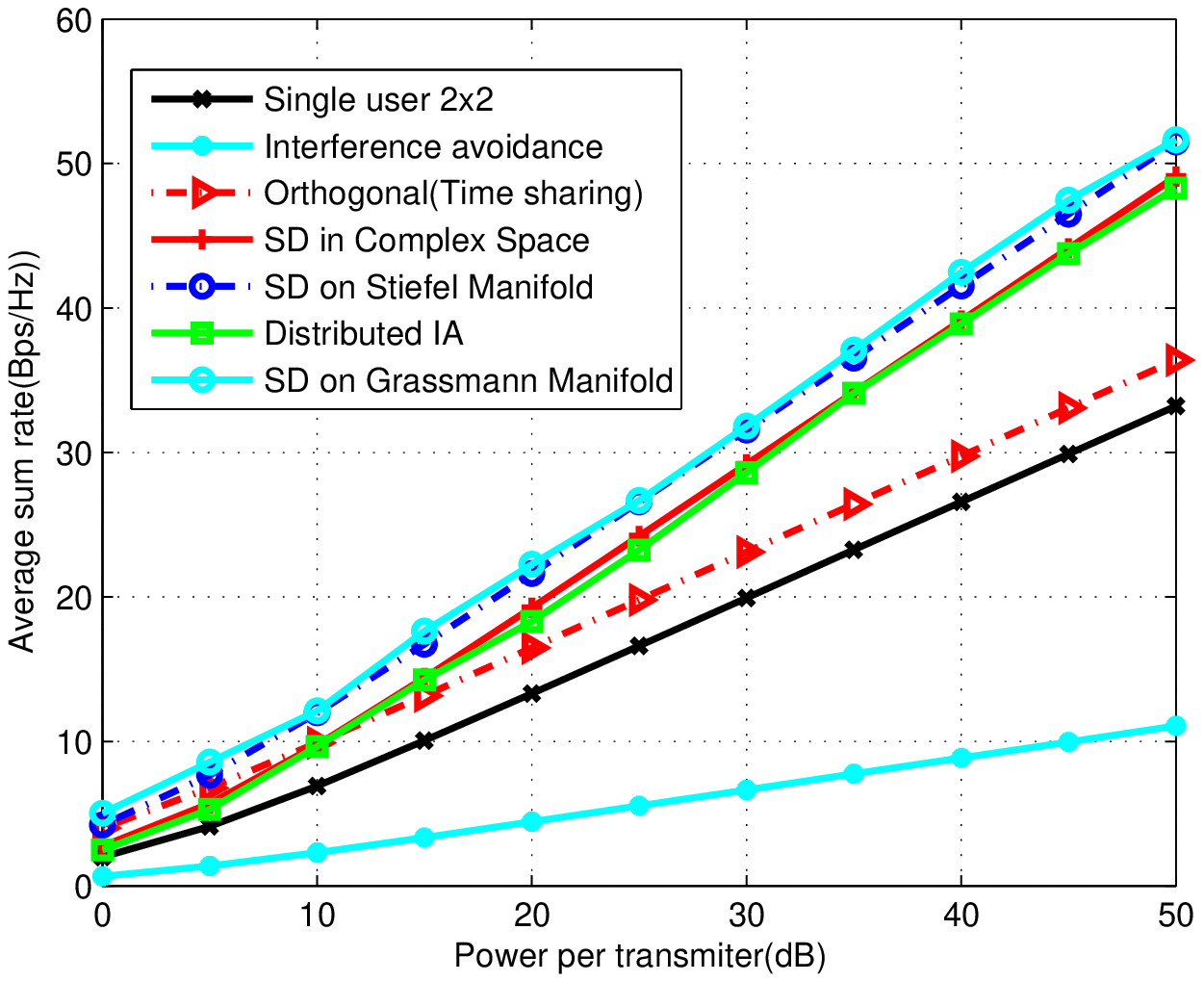}
\caption{Sum-rate capacity }
\label{sumrate}
\end{figure}

Finally we compare the system sum-rate of the proposed algorithms. Fig. \ref{sumrate} shows that the SD algorithm on Stiefel manifold and the SD algorithm on Grassmann manifold almost have the same performance, and  outperform the other classical optimization algorithms. More importantly, at high SNR the DoF of the three proposed algorithms nearly achieve 3, which is the maximum theoretical value ($KM/2 = 3$) . Therefore the perfect interference alignment is successfully achieved.

%\section{Discussion}
%\subsection{The Steepest Descent Algorithm in Complex Space }
%As previous stated, comparing with the algorithm in \cite{IEEEhowto:Ghauch}, our modified SD algorithm in complex space has lower computing complexity and higher robustness. As shown in Fig. 5, the modified SD algorithm in complex space almost has the same convergence performance as algorithm in \cite{IEEEhowto:Ghauch}. In addition, as shown in Fig. 6, the sum-rate curve of the SD algorithm in complex space and the curve of distributed IA in \cite{IEEEhowto:Jafarapp} almost overlap. While the authors in \cite{IEEEhowto:Ghauch} claimed that, the capacity performance of the work in \cite{IEEEhowto:Ghauch}  will not triumph over that of \cite{IEEEhowto:Jafarapp}.  These two evidence guarantee that the modification does not degrade the algorithm performance.
%\subsection{The Steepest Descent Algorithm on Complex Stiefel Manifold }

Two reasons leading to the fact that the  algorithms on  manifolds obtain higher system capacity  are presented below:
\begin{itemize}
 \item (i) It is noticed that our cost function actually is the interference power spilled from the interference space to the desired signal space. With the better convergence performance, the SD algorithms on  manifolds will have less remnant interference in the desired signal space within same iteration times. Therefore, the SD algorithms on  manifolds will get higher SINR\cite{IEEEhowto:Jafarapp}:
\begin{equation}
SINR = {{signal~power} \over {noise + remnant~interference}}
\end{equation}
which leads to high capacity.
\item (ii) At each receiver, the zero forcing filter is adopted. It will project the desired signal power and the remnant interference onto the subspace which is orthogonal with the subspace spanned by the interference. After performing the SD algorithms on  manifolds, it is observed that in the Euclidean norm distance,  the subspace spanned by desired signal is more close to the orthogonal complement of the  interference subspace. Therefore, even the proposed  algorithms on manifolds finally get the same remnant interference  as the classical optimization methods results. The  algorithms on  manifolds will suffer from less power lose during the projection operated by zero forcing filter,  hence higher system capacity can be achieved.
\end{itemize}

%\centering
%\includegraphics[width=3.5in]{angle1.eps}
%\caption{Angle between interfering spaces at each receiver}
%\label{p3}
%\end{figure}
%
%\begin{figure}
%\centering
%\includegraphics[width=3.5in]{angle1.eps}
%\caption{Angle between interfering spaces at each receiver}
%\label{pic3}
%\end{figure}

We notice that  better throughputs may
be attained by using non-unitary precoders, or
by applying power water-filling in the
equivalent non-interfering MIMO channels. Nevertheless,
these methods to  increase  throughputs can be performed as the
second step after the interference alignment is achieved \cite{IEEEhowto:Santa}.   
Thus  in this paper, we only need to concentrate on the first step to find the perfect solutions of interference alignment.
%and explore the second step to increase the throughputs further in the future work .
\section{Conclusion and Future Work}
In this paper,  we focus on the precoder designs on manifolds for interference alignment. By  restricting  the optimization  only at the transmitters' side, it will alleviate the significant overhead induced by alternation between the forward and reverse links. A classical SD algorithm in multi-dimensional complex space is proposed first. Then we reform the optimization problem on Stiefel manifold, and propose a novel SD algorithm on this manifold  with lower dimensions. Moreover, aiming at further reducing the complexity, Grassmann manifold is introduced to derive corresponding algorithm for reaching the perfect interference alignment.   Numerical simulations show that comparing with previous methods, the proposed  algorithms on manifolds have better convergence performance and higher system capacity, also achieve the maximum DoF. 

%Finally we prove that our algorithms converge monotonically with respect to the overall objective. 
The proof for convergence of the  proposed algorithms is quite simple. Our cost function is non-negative with the  low bound  zero. It monotonically decreases within each iteration. Therefore it must converge to a solution which is very close to zero. However, there is no guarantee that our cost function is convex\cite{IEEEhowto:Jafarapp}. Thus finding a global optimum is  in our future work.

We also do some simulations by using more sophisticated algorithm, such as Newton-type method, to achieve quadratic convergence. Yet, except for the increased computational complexity, the Newton method will converge to the closet critical point \cite{IEEEhowto:Manton}. Therefore the Newton method coupled with the steepest descent algorithm  (a few iterations of the SD algorithm are performed first to move close to a local minimum before the Newton algorithm is applied) will be investigated in our future work too.
%\begin{figure}

\section*{Acknowledgment}
This work was supported by National Natural Science Foundation of China (No. 61171112) and MIIT of China (No. 2010ZX03005-001-02).

% that's all folks
\end{document}